\documentclass[final]{aipproc}

\usepackage{epsfig}
\usepackage{amssymb}
\layoutstyle{8x11single}

\begin{document}
\title{The Abnormally Weighting Energy Hypothesis:\\ The origin of the cosmic acceleration}


\author{J.-M. Alimi}{
 address={CNRS, Laboratoire Univers et Th\'eories (LUTh), UMR 8102 CNRS, Observatoire de Paris, Universit\'e Paris Diderot ; 5 Place Jules Janssen, 92190 Meudon, France}}

\author{A. F\"uzfa}{
  address={Department of Mathematics, University of Namur (FUNDP); Rue de Bruxelles, 61, B-5000 Namur, Belgium},
  altaddress={Center for Particle Physics and Phenomenology (CP3), Universit\'e catholique de Louvain, Chemin du Cyclotron, 2, B-1348 Louvain-la-Neuve, Belgium}
}

\begin{abstract}
We generalize tensor-scalar theories of gravitation by the introduction of an abnormally weighting type of energy. This theory of tensor-scalar anomalous gravity is based on a relaxation of the weak equivalence principle that is now restricted to ordinary visible matter only. As a consequence, the convergence mechanism toward general relativity is modified and produces naturally cosmic acceleration as an inescapable gravitational feedback induced by the mass-variation of some invisible sector. The cosmological implications of this new theoretical framework are studied. This glimpses at an enticing new symmetry between the visible and invisible sectors, namely that the scalar charges of visible and invisible matter are exactly opposite.
\end{abstract}


\maketitle

\section{Introduction}
Recent observational evidence \cite{wmap1,wmap3,wmap5,perlmutter,riess,snls} indicates that the Universe is presently dominated by an intriguing component dubbed Dark Energy (DE). Its gravitational action is to drive the current cosmic acceleration by mimicking a fluid of puzzling negative pressure acting on cosmological scales. Although a huge and still unexplained fine-tuning is still required to reduce drastically the theoretical expectation of the cosmological constant value \cite{weinberg}, nevertheless, this new possible fundamental constant enters the description of the dark sector within the so-called concordance model $\Lambda\textrm{CDM}$ together with the cold dark matter CDM. 
But, this model must face a triple coincidence problem: why do we live in an almost flat Universe ($\Omega_T =1$) with roughly the same amount of baryons, DM and DE today 
($\Omega_b=0.04\approx\Omega_{DM}=0.2\approx\Omega_{DE}=0.76$)? More specifically, how could the vacuum energy be precisely of the same order of magnitude of other present cosmological components? Instead, the measured amount of DE suggests that it is ruled by some cosmological mechanism such as quintessence or generalised additional fluid components \cite{copeland} whose origin has to be found in high-energy physics. However, one can expect \cite{brax} that the difficulties encountered in trying to overcome the coincidence issues and the related problems in high-energy physics and gravitation will remain as long as DE will be regarded as an additional component \textit{independent} of baryons and DM.
\\
\\
A possible way out of the coincidence problem could then be in considering a more sophisticated physics for the whole dark sector \cite{amendola,farrar,corasaniti},
for instance by introducing in this sector new long-ranged interactions. The interesting point is then that these novel interactions in the dark sector makes \textit{only} the mass of the invisible particles varying which constitutes a violation of \textit{weak equivalence principle} (WEP). This principle has been extremely well-verified, notably at the $10^{-12}$ level with laboratory masses \cite{su}. However, these tests hold for ordinary matter \textit{only}, while the question of its validity to an invisible sector still remains an open debate, on both observational \cite{kamionkowski} and theoretical \cite{farrar,massd} points of view.
\\
\\
If the WEP does not apply to an invisible sector, then
the \textit{strong equivalence principle} (SEP), that includes the WEP and extends it to gravitational binding energies, also does not hold anymore. It is indeed clear that the gravitational energy of mass-varying invisible matter particles \textit{do not weigh} in the same way than the gravitational energy of ordinary matter particles with constant masses. Therefore, in a mixture of ordinary matter and mass-varying invisible matter, like the large-scale Universe, one should expect to observe an inescapable violation of the SEP. This crucial point has not been investigated in previous works on coupled DM-DE. In this paper, we develop a theoretical framework which shows that cosmic acceleration is precisely the observable trace of this SEP violation coming from the fact that the WEP does not apply to the invisible sector of cosmology.
\\
\\
Our basic assumption is that the invisible sector is constituted by an Abnormally Weighting type of Energy (AWE Hypothesis) \cite{awebi,awe,awedm,awedm2,awedm3}, 
i.e. the WEP does not apply to the dark sector and has therefore to be relaxed.
Doing so, one must also consider the consequent violation of the SEP as we mentionned above, meaning that the usual laws of gravitation are modified and ordinary visible matter experiences a varying gravitational strength due to the existence of the invisible gravitational outlaw. 
In such a framework, it is clear that cosmic expansion is affected and we will show that the observed cosmic acceleration can be accounted for exclusively with this mechanism,
without requiring to any explicit negative pressures. 
\\
\\
The structure of this paper is as follows.
We first build from the AWE hypothesis a tensor-scalar theory of anomalous gravity
that naturally generalises the usual tensor-scalar theories (TST) of gravitation \cite{ts,bd,convts,barrow,damour2}. Indeed, while these TST \textit{only} considered a violation of the SEP, this new framework allows to adequately describe
the modifications of gravitation that are introduced by a relaxation of the WEP, and the consequent violation of SEP. The modified convergence mechanism toward general relativity with standard ordinary matter and an AWE component are studied in the case of general equations of state.
We then relate this modified convergence mechanism to cosmic acceleration. 
We also establish there the very general conditions upon which cosmic acceleration can be achieved. We derive remarkable cosmological predictions from the analysis of Hubble diagrams of far-away supernovae that unveils the nature of the abnormally weighting component and the origin of cosmic acceleration. 
We finally  use SNe Ia data to constraint the scalar charges of this model
and show an interesting evidence for a new symmetry between the visible and hidden sectors of cosmology. 

\section{Cosmological dynamics in tensor-scalar anomalous gravity}
The \textit{Abnormally Weighting Energy} (AWE) hypothesis \cite{awebi,awe,awedm,awedm2,awedm3} suggests an interpretation of DE in terms of a relaxation of the WEP, this last being now restricted to standard ordinary visible matter.
We consider a physical framework which contains three different sectors: gravitation, the common matter (baryons, photons, etc.) and the invisible sector, made of some \textit{abnormally weighting energy}.
The laws of gravitation
have then to rule both the expansion of space-time and the variation of the gravitational strengths that are experienced differently by the two matter sectors \cite{awedm, awedm3, damour,catena}.
From the point of view of visible matter, these different gravitational strengths are not perceptible because the masses of ordinary matter particles are constants as a consequence of the restricted WEP.
However, the masses of invisible particles vary manifestly meanwhile the gravitational strength changes, yielding to cosmic acceleration as we shall see further. 
The restriction of the WEP to ordinary matter can be observed through the violation of the SEP, i.e. the changing gravitational coupling that it yields. 
This can be expressed in the observable Dicke-Jordan frame by the following action ($c=1$): 
\begin{eqnarray}
\label{s_awe}
S&=&\frac{1}{16\pi G}\int\sqrt{-\tilde{g}}d^4\tilde{x}\left\{\Phi\tilde{R}-\frac{\omega_{BD}(M(\Phi))}{\Phi}\tilde{g}^{\mu\nu}\partial_\mu\Phi\partial_\nu \Phi\right\}+S_m\left[\psi_m, \tilde{g}_{\mu\nu}\right]\nonumber\\
&&+S_{awe}\left[\psi_{awe},M^2(\Phi)\tilde{g}_{\mu\nu}\right]
\end{eqnarray}
where $G$ is the "\textit{bare}" gravitational coupling constant (reducing to Newton's constant $G_N$ on Earth), $\tilde{g}_{\mu\nu}$ is the metric coupling universally to ordinary matter, $\Phi$ is a scalar degree-of-freedom scaling the observed gravitational strength $G_c=G/ \Phi$, $\tilde{R}$ is the scalar curvature build upon $\tilde{g}_{\mu\nu}$, $\omega_{BD}(\Phi)$ is the Brans-Dicke coupling function while $\psi_{m,awe}$
are the fundamental fields entering the physical description of the matter and abnormally weighting sectors, respectively. It is assumed here that the fields of the visible and invisible sectors $\psi_m$ and $\psi_{awe}$ do not couple directly (or extremely weakly) so that a direct observation of WEP violation on matter cannot be observed from this channel in any precision test of the WEP.
In this model, the local laws of physics for ordinary matter are those of special relativity (as the matter action $S_m$ does not explicitely depend on the 
scalar field $\Phi$) while the abnormally weighting dark sector $S_{awe}$ exhibits a mass-variation (represented by the non-minimal coupling 
$M(\Phi)$). The action (\ref{s_awe}) therefore generalizes usual TST \cite{ts,bd,convts,barrow} which consider a violation of the SEP only by encompassing the whole physics of the \textit{equivalence principles} (EPs) violation due to the anomalous gravity of the dark sector. In the following, a $\tilde{}$ will denote observable quantities, i.e. quantities expressed in the Dicke-Jordan frame.
\\
\\
It is insightful 
to rewrite action (\ref{s_awe}) in the so-called
Einstein frame where the tensorial $\tilde{g}_{\mu\nu}$ and scalar $\Phi$ gravitational degrees of freedom separate into a bare metric $g_{\mu\nu}$ and a screening field $\varphi$. Moving between
the Dicke-Jordan observable frame and the Einstein frame is effected by performing the conformal transformation :
$$
\tilde{g}_{\mu\nu}=A_m^2(\varphi)g_{\mu\nu}
$$
together with the identifications:
\begin{eqnarray}
G/G_c&=&\Phi=A_m^{-2}(\varphi),\label{gc}\\
3+2\omega_{BD}(\Phi)&=&\left(\frac{d\ln A_m(\varphi)}{d\varphi}\right)^{-2}\nonumber\\
M(\Phi)&=&\frac{A_{awe}(\varphi)}{A_m(\varphi)}\cdot\nonumber
\end{eqnarray}
Doing so, the action (\ref{s_awe}) can be rewritten
\begin{eqnarray}
\label{s_awe2}
S&=&\frac{1}{16\pi G}\int\sqrt{-g}d^4x\left\{R-2g^{\mu\nu}\partial_\mu\varphi\partial_\nu\varphi\right\}+S_m\left[\psi_m,A_m^2(\varphi)g_{\mu\nu}\right]\nonumber\\
&+&S_{awe}\left[\psi_{awe},A_{awe}^2(\varphi)g_{\mu\nu}\right]\cdot
\end{eqnarray}
It should be reminded that the observable quantities are not directly obtained in this frame as the physical units are universally scaled here with $A_m(\varphi)$ (the standards
of physical units like meter and second are defined with rods and clocks that are build upon the matter fields $\psi_m$). In particular, the inertial masses of the ordinary and abnormally weighting matter sectors are scaled respectively with $A_m(\varphi)$
and $A_{awe}(\varphi)$ in the Einstein frame, while the Planck mass defines the bare gravitational strength $\bar{m}_{Pl}=G^{-2}$. The AWE hypothesis assumes that the invisible sector experiences background space-time ($g_{\mu\nu}$) with a different gravitational strength than ordinary visible matter, which is formulated
in terms of the nonuniversality of the nonminimal couplings $A_m(\varphi)\ne A_{awe}(\varphi)$, an idea echoing the effective theories of gravitation derived from string theory \cite{damour2,gasperini1,gasperini2}. This nonuniversality of gravitation is therefore a \textit{minimal} violation of the WEP taken as a whole.
\\
\\
Let us therefore study the cosmological evolution of the observed large-scale gravitational strength $G_c$ (\ref{gc}) (see also \cite{awedm,awedm3}). Assuming 
a flat Friedmann-Lema\^itre-Robertson-Walker (FLRW) background space-time $g_{\mu\nu}$ and a fluid description of the matter and AWE sectors $S_m$ and $S_{awe}$, we obtain 
the Friedmann, acceleration and scalar field equations
\begin{eqnarray}
\label{friedmann}
\left(\frac{\dot{a}}{a}\right)^2&=&\frac{\dot{\varphi}^2}{3}+\frac{8\pi G}{3}\left(\rho_m+\rho_{awe}\right)\\
\frac{\ddot{a}}{a}&=&-\frac{2}{3}\dot{\varphi}^2-\frac{4\pi G}{3}\left[\left(\rho_m+3p_m\right)+\left(\rho_{awe}+3p_{awe}\right)\right],\label{acc}\\
\ddot{\varphi}+3\frac{\dot{a}}{a}\dot{\varphi}&=&-4\pi G\left\{\alpha_m(\varphi)\left(\rho_m-3p_m\right)+\alpha_{awe}(\varphi)\left(\rho_{awe}-3p_{awe}\right)\right\}\label{kg_awe}
\end{eqnarray}
where a dot denotes a derivation with respect to the time $t$ in the Einstein frame and where $\rho_X$ and $p_X$ stand for the energy density and 
pressure of the fluid X ($\equiv$ matter or AWE) in the Einstein frame. These \textit{bare} energy densities are related to the observable ones $\tilde{\rho}_X$ by $\rho_X=A_m^4(\varphi)\tilde{\rho}_X$. Similar relation holds for the pressures.
In the Einstein frame, it is important to notice that there cannot be any acceleration unless one of the fluid violates of the strong energy condition so that $p_i<-\rho_i/3$. 
The bare expansion $a$ is then always decelerated $\ddot{a}<0$ in this frame while cosmic acceleration might occur in the observable frame under appropriate conditions on 
the scalar field dynamics. The conservation equations write
\begin{eqnarray}
\nonumber
\dot{\rho}_{i}+3\frac{\dot{a}}{a}\left(\rho_{i}+p_{i}\right)=\alpha_{i}(\varphi)\; \dot{\varphi}\left(\rho_{i}-3p_{i}\right)
\end{eqnarray}
and can be directly integrated to give
\begin{eqnarray}
\label{rhos}
\rho_{X} &=& A_{X}^{1-3\omega_{X}}(\varphi) \rho_{X,i}a^{-3(1+\omega_{X})}
\end{eqnarray}
where $\omega_{X}=p_{X}/\rho_{X}$ is the equation of state for the fluid $X$.
In the above equation, $\rho_{X,i}$ are arbitrary constants and
we will denote in the following by $R_i$ the ratio of $\rho_{m,i}$ over $\rho_{awe,i}$ at a given epoch $i$.\\
\\
Let us now reduce (\ref{kg_awe}) to a decoupled equation by using the number of e-foldings : $\lambda=\ln (a/a^i)$ as a time variable. Using (\ref{friedmann}) and (\ref{acc}), the Klein-Gordon equation (\ref{kg_awe}) now reduces to
\begin{equation}\label{dyn_1}
\frac{2\varphi''}{3-\varphi^{'2}}+(1-\omega_T)\varphi'
+\alpha_m(\varphi)(1-3\omega_T)+\frac{(\alpha_{awe}(\varphi)-\alpha_m(\varphi))}{1+\frac{\rho_m}{\rho_{awe}}}(1-3\omega_{awe})=0
\end{equation}
where a prime denotes a derivative with respect to $\lambda=\ln (a/a^i)$ and where the total equation of state $\omega_T$ for the admixture of matter and AWE fluids is $\omega_T=(p_m+p_{awe})/(\rho_m+\rho_{awe})$. The first three terms of Eq. (\ref{dyn_1}) represents the usual cosmological dynamics of TST with WEP \cite{convts,barrow} while the last brings all the information on the WEP relaxation. Obviously, the WEP can be retrieved and the well-known convergence mechanism of usual TST holds in three different cases:
(i) the nonminimal coupling is universal $\alpha_m=\alpha_{awe}$ ($M(\Phi)=1$), (ii) the AWE fluid is relativistic $\omega_{awe}=1/3$ and (iii) the AWE sector is sub-dominant $\rho_m\gg\rho_{awe}$ ($M(\Phi)\rightarrow 1$). This last case occurs for instance during the radiative era when $\rho_m\sim a^{-4}$ while $\rho_{awe}\sim a^{-3}$
for non-relativistic AWE fluid.
In conclusion, the WEP can then be well approximated anywhere the amount of invisible AWE matter is negligible.
\\
\\
When the equations of state of matter and AWE are identical $\omega_m=\omega_{awe}=\omega_T=\omega$, Eq. (\ref{dyn_1}) can easily be reduced to an autonomous equation
\begin{equation}\label{dyn_2}
\frac{2\varphi''}{3-\varphi^{'2}}+(1-\omega)\varphi'+(1-3\omega)\aleph(\varphi)=0,
\end{equation}
with $\aleph(\varphi)=d\ln\mathcal{A}(\varphi)/d\varphi$ with $\mathcal{A}(\varphi)$ the coupling function resulting from the mixing of matter and AWE fluids:
\begin{equation}
\label{biga_w}
\mathcal{A}(\varphi)=A_{awe} (\varphi)\left[R_i^{-1}+\left(\frac{A_{m}(\varphi)}{A_{awe}(\varphi)}\right)^{1-3\omega}\right]^{1/(1-3\omega)}\cdot
\end{equation}
The above equations (\ref{dyn_2})-(\ref{biga_w}) generalize
the results presented in \cite{awedm} for matter-dominated era \textit{only}.
Indeed, during the matter-dominated era ($\omega=0$), we found in \cite{awedm}
\begin{equation}
\label{biga}
\mathcal{A}(\varphi)=A_m(\varphi)+R_i^{-1}A_{awe}(\varphi),
\end{equation}
and 
\begin{equation}
\label{aleph}
\aleph(\varphi)=\alpha_m(\varphi)+\frac{(\alpha_{awe}(\varphi)-\alpha_m(\varphi))}{1+R_i\frac{A_m(\varphi)}{A_{awe}(\varphi)}}\cdot
\end{equation}
Eq. (\ref{dyn_2}) is completely
analogous to that of a damped oscillator with a variable mass rolling down some potential given by the logarithmic derivative of the resulting coupling function $\mathcal{A}(\varphi)$. Therefore, the convergence mechanism of TST with WEP is preserved despite the violation of WEP. However, this mechanism depends on the relative
concentrations of ordinary matter and AWE. In consequence, substantial variations of $G_c$ (\ref{gc}) are expected on cosmological scales where the invisible component is profuse while on very low scales they will be quite low due
to the very high density ratio of visible over invisible matter.
\section{Cosmic acceleration emerging naturally from tensor-scalar anomalous gravity}
We study here the modifications brought by the AWE component on the convergence mechanism toward general relativity (GR) and how this revised mechanism
constitutes a natural process of the observed cosmic acceleration. Indeed, this acceleration can only be described in the Dicke-Jordan frame of Eq.(\ref{s_awe}), where
cosmic expansion is described by the scale factor $\tilde{a}=A_m(\varphi)a=a/\Phi^{1/2}$ that is measured with visible matter. The Hubble diagram test allows to measure it through
the luminous distance $d_L(\tilde{z})=(1+\tilde{z})\tilde{H}_0\int_0^{\tilde{z}}
d\tilde{z}/\tilde{H}(\tilde{z})$ where the observable Hubble parameter $\tilde{H}(\tilde{z})$ is given by
\begin{equation}
\label{hubble1}
\tilde{H}^2=8\pi G A_m^2(\varphi)\left(\tilde{\rho}_m+\tilde{\rho}_{awe}\right)\frac{\left(1+\alpha_m\varphi'\right)^2}{3-\varphi^{'2}}\cdot
\end{equation}
The Friedmann equation (\ref{hubble1}) in the observable frame allows us to define the density parameters
of ordinary and AWE matter by 
\begin{equation}
\Omega_{m,\textrm{awe}}=\frac{8\pi G A_m^2(\varphi) \tilde{\rho}_{m,awe}}{(3\tilde{H^2})}
\end{equation} 
and $\Omega_\varphi=1-\Omega_m-\Omega_{\textrm{awe}}$.
The acceleration equation in the Dicke-Jordan frame is given by
\begin{eqnarray}
\label{acc_obs}
\frac{1}{\tilde{a}}\frac{d^2\tilde{a}}{d\tilde{t}^2}&=&\overbrace{\tilde{H}^2\frac{\varphi^{'2}\left(3\frac{d\alpha_m}{d\varphi}-2\right)-6\alpha_m\varphi'}{3\left(1+\alpha_m\varphi'\right)^2}}^{I}\overbrace{-\frac{4\pi G A_m^2(\varphi)}{3}\left(1+3\alpha_m^2\right)\tilde{\rho}_m}^{II}\nonumber\\
&& \underbrace{-\frac{4\pi G A_m^2(\varphi)}{3}\left(1+3\alpha_m\alpha_{awe}\right)\tilde{\rho}_{awe}\cdot}_{III}
\end{eqnarray}
In the above equation, there are only two terms that can possibly lead to cosmic acceleration: the term I, related to the dynamics of the scalar field, and the term III, in the case where the product $\alpha_{awe}\alpha_m$ would be negative. This term comprises the exchange of scalar particles between visible matter and AWE
and is actually the main contribution to cosmic acceleration (see Figure 1). 

\begin{figure}
\includegraphics[scale=0.4]{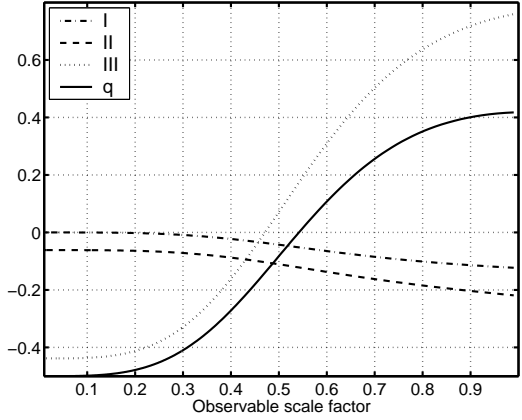}
\caption{Observable acceleration factor $q=d^2\tilde{a}/d\tilde{t}^2/(\tilde{a}\tilde{H}^2)$ obtained from the fit to SNe Ia data and its constitutive components from Eq. (\ref{acc_obs}) ($R_\infty=1$, $R_i=0.14$, $k=5.42$, $\varphi_i\approx 10^{-9}$, $\Omega_m^0=0.05$, $\Omega_{\textrm{awe}}^0=0.24$,
$\chi^2/\textrm{dof}=1.06$ on SNLS data set, $\chi^2/\textrm{dof}$($\Lambda$CDM)=$1.05$) } \label{fig1}
\end{figure}

In this figure, we have plotted separately the terms I, II and III (divided by $\tilde{H}^2$) together with their sum that
gives the observable acceleration factor $q=d^2\tilde{a}/d\tilde{t}^2/(\tilde{a}\tilde{H}^2)$ accounting for SNLS supernovae data \cite{snls}. From these data, it appears that only the term III is positive and is the main contribution to $q$ at late times. Cosmic acceleration requires $\alpha_{awe}\alpha_m<0$ or, equivalently, that
the scalar coupling strengths $\alpha_i$'s in Eq.(\ref{aleph}) have to be of opposite signs. In other words, the matter and AWE coupling functions should be inversely proportional:
\begin{equation}
\label{as}
A_{awe}(\varphi)=A_{m}^{-R_\infty}(\varphi)
\end{equation}
where $R_\infty=\rho_b/\rho_{awe}(t\rightarrow\infty)$ is the ratio at which baryons and AWE densities freeze once the scalar field reaches the attractor $\varphi_\infty$.
This parameter $R_\infty$ also measures the (absolute value of the) ratio between the opposite scalar charges of ordinary matter (baryons) and AWE related to the new interaction mediated by $\varphi$. 
\\
\\
The shape of the effective potential Eq.(\ref{biga}) during the matter-dominated era on which the scalar field rolls in Eq. (\ref{dyn_2}) looks like the shape of mexican hat illustrated in Figure 2a. It has one unstable equilibrium at $\varphi=0$ corresponding to GR with bare gravitational strength $G$ and another stable equilibrium $\varphi_\infty$ where the theory is similar to GR but with a greater value of the observed gravitational strength $G_c=G A^2_m(\varphi_\infty)$ (\ref{gc}). The gravitational coupling starts from rest at the CMB epoch near the bare value corresponding to GR, unstable in the matter-dominated era, and accelerates downward the minimum on the right.
This accelerated growth makes background space-time expansion appearing stronger and stronger, therefore reproducing DE effect. When the coupling settles into the miminum, gravitation on large-scales resembles to GR but with a stronger coupling. This mechanism depends on the relative concentration of baryons and AWE, as shown here by the values of the parameter $R_i=\rho_b/\rho_{awe}$ at CMB, and fails to depart from GR when the amount of AWE is negligible ($R_i>1$). On Figure 2b, the reader will find the modification
of the shape of the effective coupling function $\mathcal{A}(\varphi)$ Eq. (\ref{biga_w}) that follows from a variation of the equation of state parameter $\omega$ shared
by the ordinary matter and AWE fluids. Starting from $\omega=-1$ where the shape of the effective coupling function $\mathcal{A}(\varphi)$ for the admixture is close to that of the matter one $A_m(\varphi)$, the mexican hat shape progressively settles when $\omega$ increases. When the fluids are relativistic $\omega=1/3$, the effective coupling function becomes singular, as these fluids decouple from the scalar field in Eq.(\ref{dyn_2}). Finally, for ultra-relativistic fluid $\omega>1/3$, the shape of the effective coupling function $\mathcal{A}(\varphi)$ approaches the one of $A_{awe}(\varphi)$ that has no global minimum and forces a departure from GR that can only be stopped by a decrease of the equation of state. This changes in the effective coupling function are particularly promising for building an inflation mechanism
where ordinary matter and AWE are constituted, for instance, by massive scalar fields.
However, this topic goes well beyond the scope of the present work and is left for future studies.

\begin{figure}
\begin{tabular}{cc}
\includegraphics[scale=0.4]{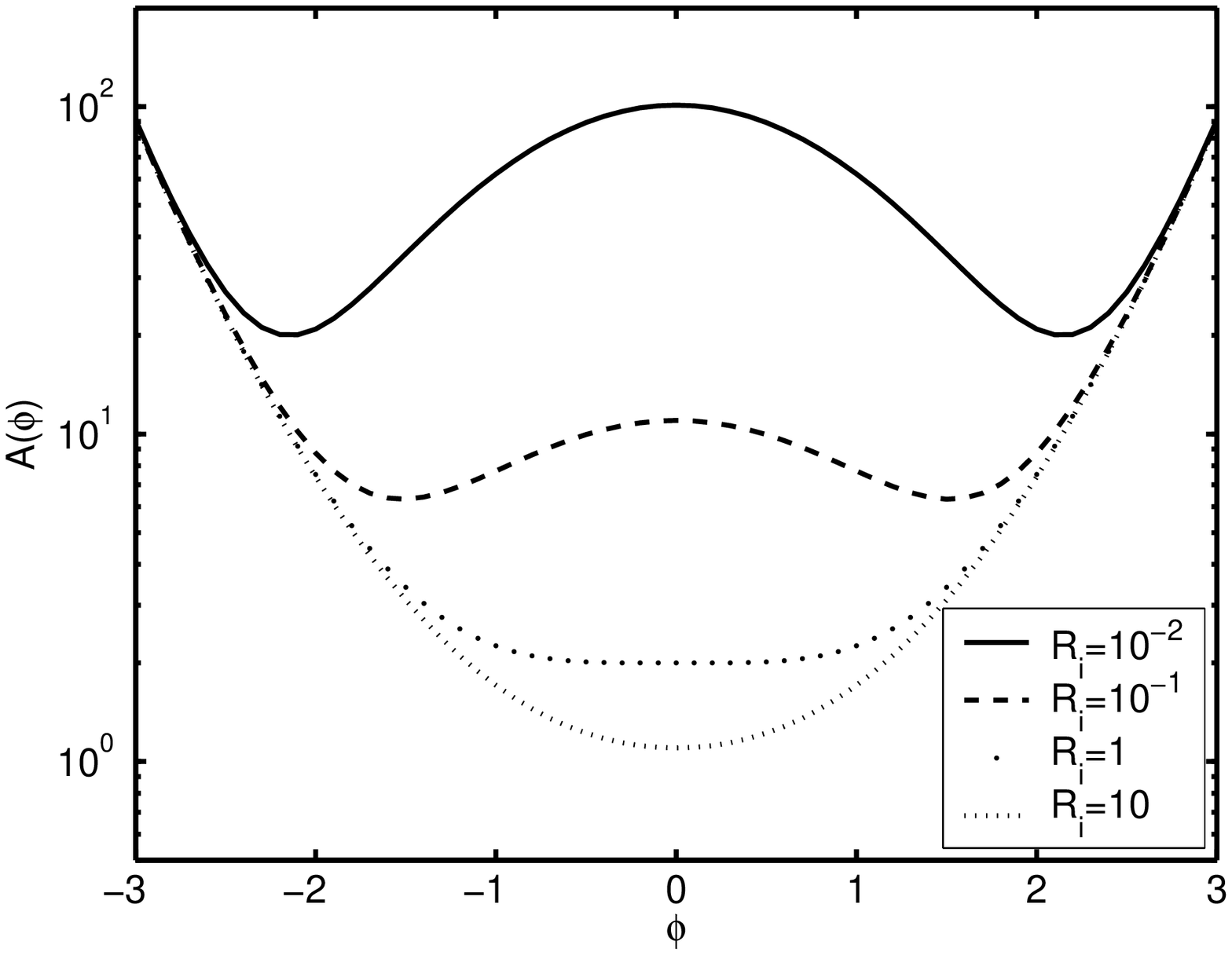} &
\includegraphics[scale=0.4]{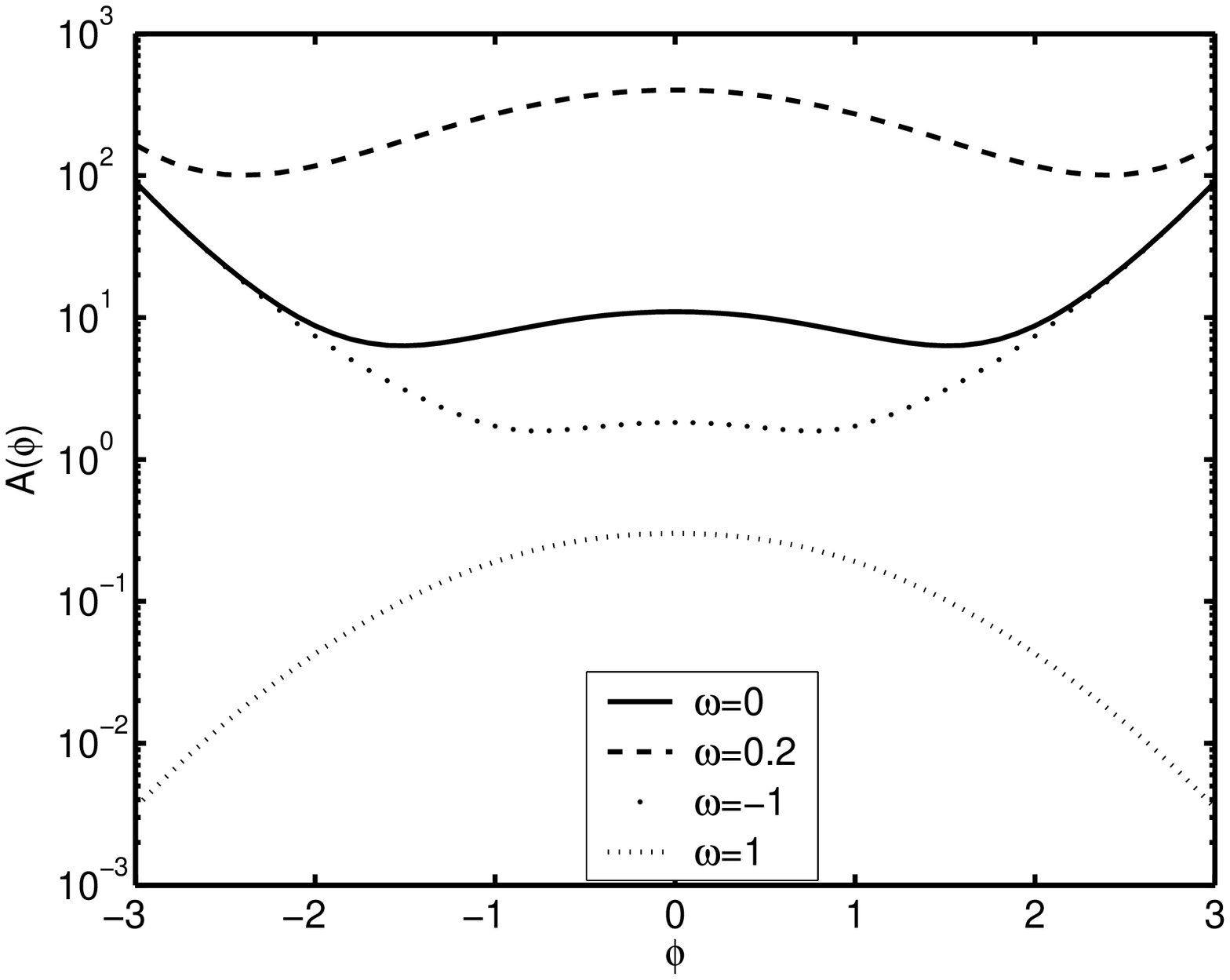} 
\end{tabular}
\caption{Effective potential for the dynamics of the gravitational coupling in the matter-dominated era (left panel) and for different
values of the equation of state parameter $\omega$ (right panel)  \label{fig2}}
\end{figure}

\section{Observational predictions of tensor-scalar anomalous gravity}
Challenging the cosmological observations with our new theoretical framework will allow us to unveil the nature of the AWE component.
To this end, we need to specify the coupling functions to be used in computing the scalar field dynamics and the cosmological observables.
From the above discussion in previous section, we need the coupling functions to AWE and matter to be reciprocal and a coupling function to matter with a global minimum.
This ensures both the existence of GR-like attractors in the modified cosmological convergence mechanism while staying compatible with tests of the SEP \cite{convts,gef}.
Without loss of generality, we can consider the following well-known parameterization for the couplings $\alpha_{m,awe} (\varphi)$ \cite{convts,awedm,awedm3,gef2,gef}:
\begin{eqnarray}
\left\{
\begin{tabular}{l}
$\alpha_m(\varphi)=k\varphi$\\
\\
$\alpha_{awe}(\varphi)=-R_\infty k\varphi$\\
\end{tabular}
\right. 
\label{couplings}
\end{eqnarray}
where $k$ is the coupling strength to the gravitational scalar. A natural assumption of nonminimally coupled theories of gravitation is to consider this strength $k$ of order unity, namely that the coupling to scalar gravitational mode $\varphi$ has the same amplitude than those of the tensorial ones $g_{\mu\nu}$.
Doing so, there are three free parameters
in this model: (i) the relative amount of matter and AWE at start $R_i=\rho_m/\rho_{awe}(a_i)$), (ii) the parameter $R_\infty$ in Eq.(\ref{as}) and (iii) the value of the scalar field at start, $\varphi_i$, which illustrates the departure from GR at the end of the radiative era.
From there, the cosmic contest between AWE and ordinary matter for ruling the EP begins and the cosmological value of $G_{c}$ is slowly pushed back from GR, provided it did not start rigorously from GR value, before falling toward the attractor whose position depends on the relative concentrations of baryons and AWE (see also Figure 2). As long as $G_{c}$ (\ref{gc}) increases, ordinary matter
couples more and more strongly to the background expansion which therefore appears accelerating to visible matter observers. \\
\\
In order to proof the adequacy of the AWE hypothesis, we have performed statistical analysis of Hubble diagram data of far-away supernovae \cite{riess,snls}.
This model predicts the following density parameters for ordinary matter and AWE as well as for the age of the Universe : $\Omega_m=0.05^{+0.13}_{-0.04}$ 
($\Omega_{m}=0.06^{+0.08}_{-0.05}$), $\Omega_{\textrm{awe}}=0.22^{+0.11}_{-0.05}$ ($\Omega_{\textrm{awe}}=0.20^{+0.08}_{-0.06}$), 
($\Omega_{\varphi}=1-\Omega_{m}-\Omega_{\textrm{awe}}$)
and $t_0=13.1^{+0.6}_{-0.4}Gyr$ ($t_0=14.1^{+1.02}_{-0.55}Gyr$) at the 95\% confidence level for HST (SNLS) data sample ($H_0=70km/s/Mpc$ for the estimation of $t_0$). 
These results strongly suggest to identify ordinary matter to baryons, AWE to DM and the dynamics of the scalar field $\varphi$ to a DE component.
Doing so, the predicted amount of baryons is consistent with Big Bang Nucleosynthesis (BBN) and CMB constraints \cite{wmap1}.
These cosmological predictions look remarkably similar to the predictions of the concordance model $\Lambda$\textrm{CDM}
although they are obtained from the Hubble diagram \textit{alone}, i.e. from the homogeneous dynamics of the Universe, which 
is an outstanding achievement of the AWE hypothesis. 
Indeed, the $\Lambda\textrm{CDM}$ is only able to predict the total amount of pressureless matter 
($\Omega_{b}+\Omega_{DM}$) from the same cosmological test and requires a cross-analysis, for instance of the CMB anisotropies, to break this degeneracy. 
In the present case, the Hubble diagram becomes a fully independent test whose results are, in addition, consistent with WMAP constraints \cite{wmap1,wmap3}. Figure 3 illustrates the confidence regions in the plane ($\Omega_{b}$, $\Omega_{awe}\equiv\Omega_{DM}$) with the parametrization of the previous coupling function. As well, we can measure, with supernovae data alone, the relative amount of baryons and AWE (i.e., DM according to the above identifications) at the beginning of the matter-dominated era (represented by the ratio of their densities at that time $R_i=\rho_b/\rho_{DM}$(CMB)).
We find $R_i=0.21^{+0.09}_{-0.07}$ ($R_i=0.22^{+0.13}_{-0.07}$) at the 68\% confidence level for HST (SNLS) data sample. 

\begin{figure}
\begin{tabular}{cc}
\includegraphics[scale=0.3]{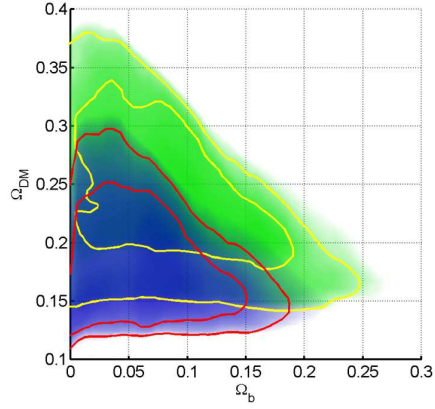} &
\end{tabular}
\caption{68\% (inner) and 95\% (outer) confidence contours of the density parameters associated to ordinary and AWE (dark matter) ($\Omega_b$ and $\Omega_{DM}$, respectively) 
for HST (green) and SNLS (blue) supernovae data sets.  \label{fig3}}
\end{figure}

In Figure 4,
we give a comparison between the observable scale factors accounting for supernovae data predicted by the present model specified by (\ref{hubble1}), (\ref{acc_obs}) and (\ref{couplings}) and the concordance model. Both cosmic expansions remains almost undistinguishable until the Universe reaches about twice its present size. Then, while the cosmological constant endlessly dominates
the Universe and drives an eternal exponential expansion, the AWE hypothesis suggests that baryons and AWE enter the final stage of their contest to rule the EP. This ultimate episode begins with the fall of the gravitational coupling into the attractor of Figure 2 which makes the scale factor oscillating. An equilibrium asymptotically establishes where baryons and DM do equally weigh (for $R_\infty =1$) and gravitation on large-scales is correctly described by GR with a different value of the coupling.

\begin{figure} 
\includegraphics[scale=0.4]{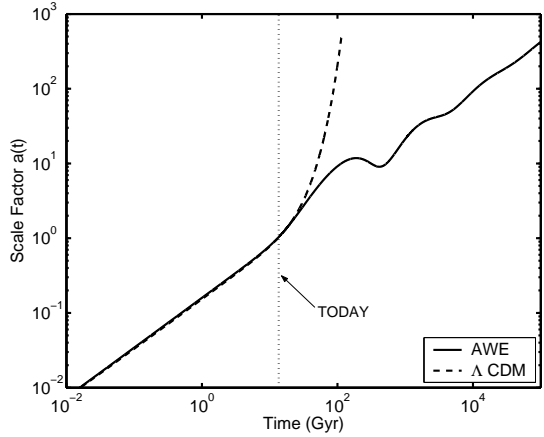}
\caption{Comparison between the cosmic histories predicted by the AWE model in the Dicke-Jordan frame and the concordance model $\Lambda\textrm{CDM}$.\label{fig4}}
\end{figure}
\section{A Mach-Dirac Principle for gravitation}
The mass-variation of the AWE, which consists on a violation of WEP between visible and invisible sectors, 
implies a compulsory violation of the SEP. This reasoning on the equivalence principles links together two old but crucial ideas in gravitational physics: Mach's principle and Dirac's hypothesis. Mach's principle \cite{bd} was one of Einstein's main inspiration \cite{pais} when he conceived GR and posits that inertia and acceleration are relative and should therefore be refered to distant matter. In particular, Mach advanced that inertial masses could only be defined with respect to the entire matter distribution in the Universe. The other key idea is the hypothesis of varying observed gravitational strength $G_c$ (\ref{gc}) that Dirac formulated \cite{dirac} in order to justify some puzzling numerical coincidences between cosmological and quantum numbers. Here we show that the AWE hypothesis can merge into a \textit{Mach-Dirac principle} these thoughtful ideas by identifying the machian mass variation of AWE as the source of Dirac's varying gravitational strength. \\
\\
Indeed, a new remarkable feature of this model is that, besides of its cosmological predictions, it gives a hint on a new relation between microphysics and gravitation.
This is expressed in terms of the gravitational strength $G_{c}$ (\ref{gc}) and DM mass $m_{{DM}}\equiv \bar{m}_{DM} M(\Phi)$, where $\Phi(x^\mu)$ is the dimensionless space-time dependent
coupling, $G_{N}$ is Newton's constant and $\bar{m}_{DM}$ is the DM mass that would be measured in Earth-based laboratories.
To obtain the cosmological evolution described above (see also Figure 2), we have seen that the gravitational coupling $G_{c}$ and the DM mass-scaling must be inversely proportional (see Eq.(\ref{as}) and \cite{awedm,awedm3}), which also means that ordinary matter and AWE must have opposite scalar charges.
In terms of the dimensionless coupling $\Phi$, we therefore have:
\begin{equation}
M(\Phi)=\Phi^{\frac{1+R_\infty}{2}}
\end{equation}
where $R_\infty$ is also the ratio $\rho_{m}/\rho_{awe}(t\rightarrow\infty)$ at which baryons and DM densities freeze into the above-mentionned attractor (see Figure 2).
From the cosmological data on supernovae, we find $R_\infty=1.23^{+0.96}_{-0.67}$ ($R_\infty=1.35^{+1.11}_{-0.85}$) at the 68\% confidence level for HST (SNLS) data sample. The measured value of $R_\infty$ close to unity indicates that the scalar charges of ordinary matter and DM are \textit{exactly} opposite. 
Indeed, $R_\infty=1$ implies that $\alpha_m(\varphi)=-\alpha_{awe}(\varphi)$ ($A_m(\varphi)=A_{awe}^{-1}(\varphi)$ ; $M(\Phi)=\Phi$) which constitutes
a hint to a possible new symmetry between the hidden and visible sectors. In other words, this value of $R_\infty$ also points to an intriguing relation between the constant mass of baryons $m_b$ and the changing DM $m_{DM}$ and observed gravitational strength $G_{c}$ (\ref{gc}):
\begin{equation}
G_c(x^\mu)\; m_{b}\;m_{DM}(x^\mu)=G_{N}\; m_{b}\;\bar{m}_{DM},
\label{md}
\end{equation}
where the bar means the Earth laboratory value. 
This dimensionless relation is frame-independent and imposing $R_\infty=1$ is the only way of constructing a constant ratio
between the masses of the particles from the visible and invisible sectors and the Planck scale. Although this relation does not fix the bare mass of DM, it rules its scaling by imposing a conservation of the product of the gravitational \textit{charges} of baryons and DM. 
This important phenomenological law, directly deduced from cosmological data and linking together gravitational scales and masses of visible and invisible matter glimpses at the intimate nature of gravitation. The deep meaning of Eq.(\ref{md}) constitutes a crucial question for the many fundamental approaches that aim to unify gravity and microphysics with
explicit space-time dependancies of masses and couplings \cite{ashtekar,connes,gasperini1,gasperini2}.
\section{Conclusion}
Arising from a critical discussion on the equivalence principles, the AWE Hypothesis leads to a new elegant theory of gravitation
for dealing with the invisible sector, \textit{tensor-scalar anomalous gravity}, that glimpses beyond GR.
Applying this theory to cosmology allowed us to successfully explains \textit{cosmic acceleration} in terms of the anomalous gravity arising as a feedback of \textit{dark matter} mass variation. Doing so, this suppresses the need of an independent additional DE component with an almost intractable coincidental dominance.
But this interpretation of DE also offers to physicists new observational and theoretical challenges.
The existence of an abnormally weighting type of energy, that one can identify to DM from the results presented here, not only affects the background expansion, at the opposite of minimally coupled models of DE, it also modifies gravitational physics on large-scales where DM is profuse. 
This offers new perspectives to test the nature of DM, and indirectly of DE itself. 
The theoretical challenge lies in explaining why the gravitational energy scale, i.e. the Planck mass, would change
with the inertial mass of DM particles, as dictated by Eq. (\ref{md}). This Mach-Dirac principle is intimately linked to an enticing new symmetry between the visible and invisible sectors: the exact oppositeness of their scalar charges. The existence of such a long-ranged fifth force mediated by a scalar exchanged between the opposite charges of the visible and invisible sectors, can be hoped to be established in local experiments \cite{kaloper,chameleon2}. 
If the AWE hypothesis really constitutes the \textit{missing link} between dark matter and dark energy, then elucidating the intimate nature of gravitation could be closer than ever. 

\bibliographystyle{aipproc} 

\end{document}